\documentclass[twocolumn,floats,a4paper,10pt]{article}

\usepackage[utf8]{inputenc}
\usepackage[USenglish]{babel}
\usepackage[]{graphicx}
\usepackage[affil-it]{authblk}
\usepackage{lipsum}
\usepackage[symbol]{footmisc}
\usepackage{hyperref}

\title{The Role of Dispersal in Competition Success and in the Emerging Diversity}
\author[1]{E. Heinsalu\thanks{\emph{e-mail:} els.heinsalu@kbfi.ee}}
\author[1,2]{D. {Navidad Maeso}}
\author[1]{M. Patriarca}
\affil[1]{\small{National Institute of Chemical Physics and Biophysics - R{\"a}vala 10, Tallinn 15042, Estonia}}
\affil[2]{\small{Tallinn University, School of Natural Sciences and Health - Narva 29, 10120 Tallinn, Estonia}}

\begin{document}

 \twocolumn[
  \begin{@twocolumnfalse}

   \maketitle
    \begin{abstract}
    
    The dynamics of dispersal-structured populations, consisting of competing individuals that are characterized by different diffusion 
    coefficients but are otherwise identical, is investigated. 
Competition is taken into account through demographic processes.
The problem addressed models natural selection.
It is observed that the mean value and the relative width of the initial distribution of the diffusion coefficients characterizing the 
individuals together with the temporal fluctuations determine the final distribution of the diffusivities (diffusion coefficients leading to 
the competition success) as well as the final diversity of the system at finite time (the number of different diffusion coefficients present 
in the system).
Large initial mean diffusivity of the system leads to a rather fast disappearance of the diversity. 
Instead, small initial mean diffusivity of the system leads to a diversity equal to the number of niches forming in the system due to the 
competitive interactions.
The cluster formation is also associated to the competition success of the slower diffusing individuals.
The diversity is diminished by the increase of the temporal fluctuations that give the competition advantage to the faster diffusing 
individuals.
Somewhat counterintuitively, under certain conditions the competition success is given by intermediate values of the diffusion 
coefficients.\\



\end{abstract}
    \strut 
  \end{@twocolumnfalse}
]

\footnotetext[1]{\emph{e-mail:} els.heinsalu@kbfi.ee}


\section{Introduction}


Observing nature, one sees that representatives of different species move in very different ways and with very different speeds. 
The variances are huge not only between animal classes, but also among species in the same class. 
Because in the nature there is typically a good reason for everything, all the various ways to move certainly have different evolutionary 
causes behind, the overall reasons being the survival and competition advantage.
In fact, the motion of organisms is a crucial component in ecological and evolutionary processes and one of the fundamental features of life, 
together with the reproduction and death processes.
Importantly, besides the ways \textit{how} to move, also the reasons \textit{why} to move are rather different.
On the individual level, the motives can be, for example, foraging, escaping from the predation, diminishing competition, search for a mate or
a suitable habitat.
On the population level, the motives can be most commonly biological dispersal or migration.
To study the animal migration and swarming is an extremely challenging task: 
to understand the dynamics of the collective behavior as well as the behavior of the individuals and also the causes behind the migration and 
swarming.
In some cases the reasons are rather clear and concern usually food, breeding, and climate; in other cases, as, for example, the migration of 
monarch butterflies, no clear explanation has been found yet.
All this --- the rich variety of movement modes seen in nature and its importance --- has lead to the rise of the field called 
\textit{movement ecology} \cite{Holden-2006}.

The object of the present paper is not to examine any concrete species, analyze the dynamics of any specific manner to move, or the causes 
for the movement events.
Instead, our aim is to study the emerging effects due to a heterogeneity in the diffusion properties of individuals, from a general ecological
perspective.
Through a stochastic individual-based model, we investigate the problem of competing individuals who diffuse with different speeds, but are 
identical in all the rest, i.e., we study {\it the dynamics of dispersal-structured populations}.
The motivation for such study comes from the fact that the dispersal ability can vary as much within a species as among species, as discussed 
in Ref.~\cite{Stevens-2010}.
The issues that we address concern the competition outcome in dispersal-structured populations, i.e., we focus on the {\it natural selection 
process} instead of the evolution of dispersal through mutations, as investigated in numerous works, 
e.g., Refs.~\cite{Novak-2014,Pigolotti-2014-PRL,Dockery-1998,Johnson-1990,Heino_Hanski_2001}, and the resulting system diversity 
(number of different diffusivities present in the system).

The work presented in the current paper is a continuation of Ref.~\cite{Heinsalu-2013-PRL} (see also Ref.~\cite{EHG-2015}) where the 
competition between Brownian and L\'evy bugs was investigated (see also Ref.~\cite{Heinsalu-2012-PRE}). 
In that paper, the conclusion was that typically the species winning the competition is the species forming stronger clusters, which in the 
case of the organisms using the same motion means the less motile species.
However, this is true only in the mean-field approximation, i.e., in the case of small temporal fluctuations; for large temporal 
fluctuations (not described by the mean-field theory) the result is the opposite \cite{Pigolotti-2014-PRL}.
Furthermore, in Ref.~\cite{Heinsalu-2013-PRL}, it was also observed that, in a certain range of diffusivities, coexistence of Brownian 
and L\'evy bugs can occur.

The coexistence of the species in the case of clustering is not captured in the mean-field theory, which predicts that the slower species 
forming stronger clusters should always win the competition. 
However, the conditions for the coexistence are not difficult to understand.
Namely, it takes place when Brownian walkers form very strong clusters that the L\'evy walkers are not able to invade despite the fact that 
they are able to wander around them. 
On the other hand, due to the extremely low diffusion and the high death rate in the inter-cluster space, the Brownian walkers are not 
capable to colonize the territories that have been occupied by the L\'evy walkers during the initial cluster formation due to random 
fluctuations \cite{Heinsalu-2013-PRL,EHG-2015}.
This observation has given the motivation for the present paper: it seems natural to expect that something similar takes place when there 
are many different diffusivities in the system and that there might be an optimal range of diffusivities leading to the competition advantage.

The structure of the paper is the following: in Sec.~\ref{Sec-model} we present the model,
in Sec.~\ref{Sec-results} we present the results;
in Sec.~\ref{Sec-conclusion} the conclusions are drawn.
Section~\ref{Sec-results}, in turn, is divided in 6 subsections.
First, we discuss the patch formation and the influence of the heterogeneity as well as temporal fluctuations on it and therefore also on 
the system diversity.
In Sec.~\ref{Sec-competition} we study the outcome of the competition between the individuals with different diffusivities.
We observe that under certain conditions the competition success is given for the individuals characterized by an intermediate diffusion 
coefficient.
In order to understand the results obtained, we investigate in Secs.~\ref{Sec-evolution} and \ref{Sec-RT} the evolution of the system in 
time as well as the residence times in states with different diversities.
Finally, in Sec.~\ref{Sec-1D} we investigate a one-dimensional system that allows to make a comparison between an analytical calculation 
and numerical result, in order to understand the underlying mechanisms of the stabilizing selection observed in Sec.~\ref{Sec-competition}.


\section{Model} \label{Sec-model}


We study by numerical simulations a system consisting of organisms that reproduce asexually, die, and move in space according to Brownian 
diffusion
({the {\it Brownian bug model}, see Refs.~\cite{Zhang-1990,Young-2001,Felsenstein-1975,Heinsalu-2012-PRE}).
We assume that initially the system consists of $N_0 = 5000$ organisms (bugs), which is much more than the carrying capacity of the system.
The bugs are placed randomly in a two-dimensional $L \times L$ square domain with periodic boundary conditions.
We assume $L = 1$, so that lengths are measured in units of system size.

The variety in the individual diffusivities is assigned through the initial conditions.
This corresponds to the situation observed in real systems, where individuals are to a greater or lesser extent all different from each other, 
due to the natural variation and mutations. 
We assume that all individuals present at time $t = 0$ are characterized by different diffusion coefficients $\kappa_j$, 
with $j = 1, \dots, N_0$, extracted randomly from a uniform distribution in the interval $[\kappa(1-d), \kappa(1+d)]$, with mean value 
$\kappa$ and standard deviation
$\sigma_{\kappa} =  \ \kappa d/\sqrt{3}$;
the parameter $d \in [0, 1]$ provides the relative width of the distribution and characterizes the initial relative heterogeneity of the 
system.
Thus, the larger is $\kappa$ and the larger is $d$ for the given $\kappa$, the more different are the diffusivities of the  individuals in 
the system, ranging across the whole interval $[0, 2 \kappa]$ for $d = 1$.
For $d = 0$, instead, all the bugs have the same diffusion coefficient $\kappa$, i.e., we recover the case of a homogeneous 
system~\cite{EHG-2004,EHG-2005,CL-2004,Heinsalu-2012-PRE,EHG-2015}.
In general, it would perhaps be more realistic to assume that the diffusivities of the individuals follow the normal distribution; however, 
we have checked that using the normal distribution instead of the uniform one does not influence the results significantly. 
Therefore, for the sake of simplicity, in the following the uniform distribution is used.

The demographic processes are affected by the competitive interactions. 
Namely, the bug labeled $i$ ($i = 1, \dots, N$, with $N \equiv N(t)$ being the number of bugs in the system at time $t$) reproduces and dies 
following Poisson processes with rates $r_b^i$ and $r_d^i$ (probabilities per unit of time), respectively, given by \cite{EHG-2004}
\begin{eqnarray}
r_b^i &=& \mathrm{max}(0,r_{b0} - \alpha N_R^i) \, ,  \label{rates-b} \\
r_d^i &=& r_{d0} + \beta N_R^i \, . \label{rates-d}
\end{eqnarray}
%
%
%
Here, $r_{b0}$ and $r_{d0}$ are the constant reproduction and death rates of an isolated bug.
The terms containing the positive parameters $\alpha$ and $\beta$ take into account the competitive interactions between the individuals:
the reproduction rate of an individual $i$ decreases and the death rate increases with the number of its neighbors $N_R^i$ that are at a 
distance smaller than $R$ from it (it is assumed that $R \ll L$).
Thus, the parameters $\alpha$, $\beta$ determine how the birth and death rates of the organisms depend on the density (on the competition on 
the resources), respectively.
Finally, the max() function in the first equation ensures the positivity of the birth rates.
The critical number of neighbors, $N_R^*$, for which death and reproduction are equally probable for bug $i$, is determined by
\begin{equation}
\label{NR*}
N_R^* = \frac{r_{b0} - r_{d0}}{\alpha + \beta}  \equiv \frac{\Delta_0}{\gamma} \, .
\end{equation}
where $\Delta_0 = r_{b0} - r_{d0}$, while $\gamma = \alpha + \beta$ is referred to as the {\it competition intensity}.
For $N_R^i < N_R^*$ it is more probable that bug $i$ reproduces and for $N_R^i > N_R^*$ death becomes more likely. 
In the case of reproduction, the newborns are placed at the same positions as the parents, leading to reproductive correlations, and they 
will inherit also their characteristics of the parents (i.e., the diffusion coefficient).

The time evolution of the system is simulated through the Gillespie algorithm and the spatial motion of the individuals is modeled as a 
two-dimensional continuous time random walk, as described in Ref.~\cite{Heinsalu-2012-PRE}, with the variance that now the individuals have 
different diffusivities and the newborns inherit the diffusion coefficients of their parents.

Throughout the article it is assumed that the interaction range is $R = 0.1$, the reproduction and death rates of an isolated bug 
$r_{b0} = 1$ and $r_{d0} = 0.1$, respectively, and that the competition intensity is $\gamma = 0.02$.
Consequently, also the critical number of neighbors is constant, $N_R^* = 45$.
Thus, the only parameters that we vary are the mean diffusion coefficient $\kappa$, the parameter $d$, that determines the heterogeneity of 
the system for given $\kappa$, and $\beta$, which determines temporal fluctuations by modulating the death rates.


\section{Results} \label{Sec-results}



\subsection{Patch formation} \label{Sec-clustering}



\subsubsection{Unstructured populations} \label{unstructured}


In the case of interacting Brownian bugs moving all with the same diffusion coefficient $\kappa$ (i.e., $d = 0$), for small enough 
$\kappa$ and large enough $\Delta _0$, the salient property of the model is the natural formation of a clumped spatial distribution 
\cite{EHG-2004,CL-2004,EHG-2015}, which is the most common distribution observed in nature.
In particular, in the case of a two-dimensional system, in the statistically steady state (small temporal fluctuations, meaning small 
$r_{d0}$ and $\beta \ll \gamma$) clusters arrange in a hexagonal lattice with the periodicity of the order of the interaction radius 
$R$ \cite{EHG-2004,EHG-2005}.
The appearance of the periodic clustering is well captured in the mean-field approach \cite{EHG-2004,EHG-2015}.
It emerges as the interplay between the reproductive correlations, limited ability of the offsprings to move away from their habitats, 
and, foremost, the competitive interactions between the individuals (see also Ref.~\cite{Heinsalu-2010-EPL}).
The mathematical condition from the mean-field theory for the pattern formation is:
\begin{equation}
\label{cond-p}
2 R^2 \Delta _0 / \kappa > \nu_c ~~ \mathrm{with} ~~ \nu_c ~ \approx ~ 370.384 \, . 
\end{equation}
Thus, for the given parameters ($R=0.1$ and $\Delta_0 = 0.9$) the diffusion coefficient has to satisfy the condition 
$\kappa < 4.86 \times 10^{-5}$.
The smaller the diffusion coefficient, the smaller is the linear size of the clusters.
Increasing the diffusion coefficient, the clusters forming the pattern become more spread and for large values of 
$\kappa$ (so that Eq.~(\ref{cond-p}) does not hold anymore) the periodic pattern is replaced by an almost homogeneous distribution of 
bugs \cite{Heinsalu-2012-PRE}.

Increasing the temporal fluctuations, i.e., increasing $\beta$ or $r_{d0}$ in Eq.~(\ref{rates-d}) keeping at the same time $\Delta_0$ 
and $\gamma$ constant (remember that $\gamma$ expresses the competition intensity), the clusters become narrower and arrange in a more 
disordered way \cite{EHG-2015,Heinsalu-2012-PRE}: the ideal periodic pattern becomes more similar to real patterns observed in nature.
The center of masses of the clusters will rather perform a random walk instead of fluctuating slightly around the fixed positions of the
periodic pattern and will occasionally even disappear from the system.
In this regime, the mean-field approximation does not describe the system anymore.
In the following we study the case of rather small temporal fluctuations (i.e., $\beta \ll \gamma$), unless indicated differently.


\subsubsection{Influence of heterogeneity} \label{inf_het}


Introducing the heterogeneity in the diffusion coefficients of the bugs, i.e., investigating a dispersal-structured population, three 
scenarios can arise in the case of sufficiently small temporal fluctuations.

For small values of the initial mean diffusion coefficient of the system, $\kappa < R^2 \Delta_0/\nu_c$ (for $R=0.1$ and $\Delta_0 =0.9$, 
thus $\kappa < 2.429 \times 10^{-5}$),  
the clustering takes place for any value of $d$, because in this case each value of $\kappa_j$ in the interval $[\kappa(1-d), \kappa(1+d)]$ 
with $d \in [0, 1]$ satisfies condition (\ref{cond-p}).

For large values of $\kappa$ no clustering appears for any value of $d$.
If $d$ is small, $d < 1 - 2R^2 \Delta_0/(\kappa \nu_c)$, the reason lies in the fact that none of the values of  $\kappa_j$ in the interval 
$[\kappa(1-d), \kappa(1+d)]$ leads to the cluster formation in the case of a homogeneous system.
For large values of $d$, $d > 1- 2R^2 \Delta_0/(\nu_c \kappa)$, so that the initial system contains also individuals characterized by small 
diffusivities leading to the clustering, the success of the cluster forming individuals is prevailed by the much larger fraction of 
individuals with larger $\kappa_j$ that create a well-mixed environment.

Instead, for a certain range of $\kappa$, the clustering of particles and pattern formation depends on the value of $d$: 
for $d > 1- 2R^2 \Delta_0/(\nu_c \kappa)$ the pattern formation can take place.
From Fig.~\ref{clustering} one can see that the system with the initial mean diffusivity $\kappa = 2 \times10^{-4}$ for which no periodic 
clustering is observed in the homogeneous case (Fig.~\ref{clustering}a), can develop the periodic pattern due to the heterogeneity in 
diffusivities, but only for sufficiently large values of $d$ (Figs.~\ref{clustering}c and \ref{clustering}d).
The mathematical limit for the given parameters is $d > 0.757$; however, the transition to pattern formation is not sharp and also smaller 
values of  $d$ lead to a certain level of clustering.
For too small values of $d$ the distribution of organisms remains similar to the one appearing in the case of unstructured populations 
(c.f. Figs.~\ref{clustering}a and \ref{clustering}b).
The comparison between Figs.~\ref{clustering}c and \ref{clustering}d reveals that clustering is the stronger the larger is $d$, i.e., the 
larger is the variation in the initial population and, consequently, 
the smaller are the diffusion coefficients of the slowest individuals, see also Secs.~\ref{unstructured} and \ref{Sec-evolution}.
Thus, the patchiness in the systems with dispersal-structured populations with equal initial mean diffusivity can be influenced by the
heterogeneity of the population (standard deviation of the initial distribution of individuals diffusivities).

\begin{figure}[!t]
\resizebox{0.45\textwidth}{!}{\includegraphics{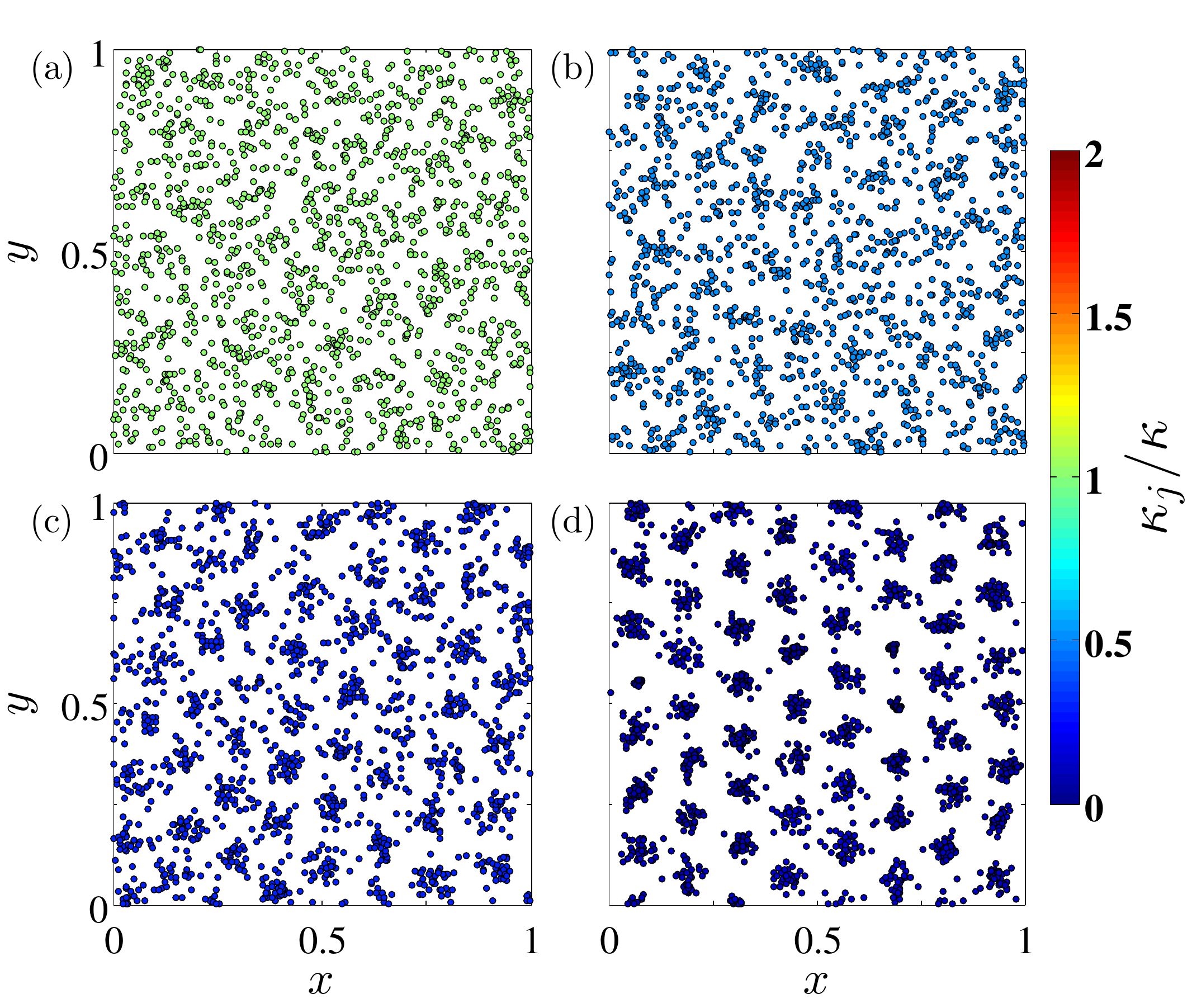}}
\caption{Appearance of the periodic pattern in the spatial distribution of the organisms due to the heterogeneity in diffusion coefficients.
The value $\kappa = 2 \times 10^{-4}$ is used for all the panels, but the heterogeneity of the systems, characterized by the parameter $d$, 
is different:
(a) $d=0$, corresponding to the homogeneous system; (b) $d=0.5$; (c) $d=0.7$; (d) $d=1.0$. The color of the bugs corresponds to their 
diffusion coefficients as indicated by the legend; $\beta = 0$.}
\label{clustering}
\end{figure}
%


\subsection{The influence of $\kappa$ and $d$ on diversity}


The formation of patches is known to be one of the key promoters for species diversity.
In fact, we observe that the systems with small value of $\kappa$, e.g., $\kappa = 10^{-5}$, where strong clustering occurs for any value 
of $\kappa_j$, leading to small probability of successful inter-cluster traveling, are characterized by a diversity $D$ (number of different 
diffusion coefficients present in the system) equal (approximately) to the number of niches, $N_c$, i.e., limited ability to move and the 
resulting isolation promotes the diversity. 
Instead, the weak clustering or almost homogeneous distribution of organisms leads to the rapid decrease of the diversity until the total 
population will consist of the individuals moving all with the same diffusivity, being thus the successors of the same ancestor, see also 
Ref.~\cite{Heinsalu-2012-PRE} and Fig.~\ref{evolution}.
The dependence of the diversity $D$ on the parameters $d$ and $\kappa$ is summarized in Fig.~\ref{phasediagram}, presenting the contour-plot 
of the surface $\langle D(\kappa, d) \rangle$ at a finite time $t = 5 \times 10^5$, where the average was taken over at least 20 simulations.
As one can see, the value of $D$ depends more strongly on $\kappa$ and is the larger the smaller the initial mean diffusion coefficient of 
the system is.
Larger values of $d$, having in general a positive effect on the clustering, as discussed in Sec.~\ref{inf_het}, lead to the increase of the 
diversity at a finite time.

Notice that due to the stochastic fluctuations in the number of individuals and the irreversibility of death, the number of different 
diffusivities decreases in time, reaching, in principle, the value $1$ after a sufficiently long time for any values of the parameters.
The required time for this is the larger, the smaller is $\kappa$ (see also Secs.~\ref{Sec-evolution} and \ref{Sec-RT}).
However, because in real systems the time is always finite (and especially, keeping in mind that we investigate natural selection, which 
is a process taking place in a limited time interval smaller than the mutation time scale), we have set a maximum simulation time 
$t_\mathrm{max} = 5 \times 10^{5}$.

On the basis of Fig.~\ref{phasediagram}, one can also define the ranges of small, intermediate, and large values of $\kappa$, leading 
(in the case of small temporal fluctuations) during the simulation time to
a $D \approx N_c$ for any value of $d$;
$1 \le D < N_c$, depending on $d$;
and to $D = 1$, for any value of $d$, respectively.

\begin{figure}[!t]
\resizebox{0.45\textwidth}{!}{\includegraphics{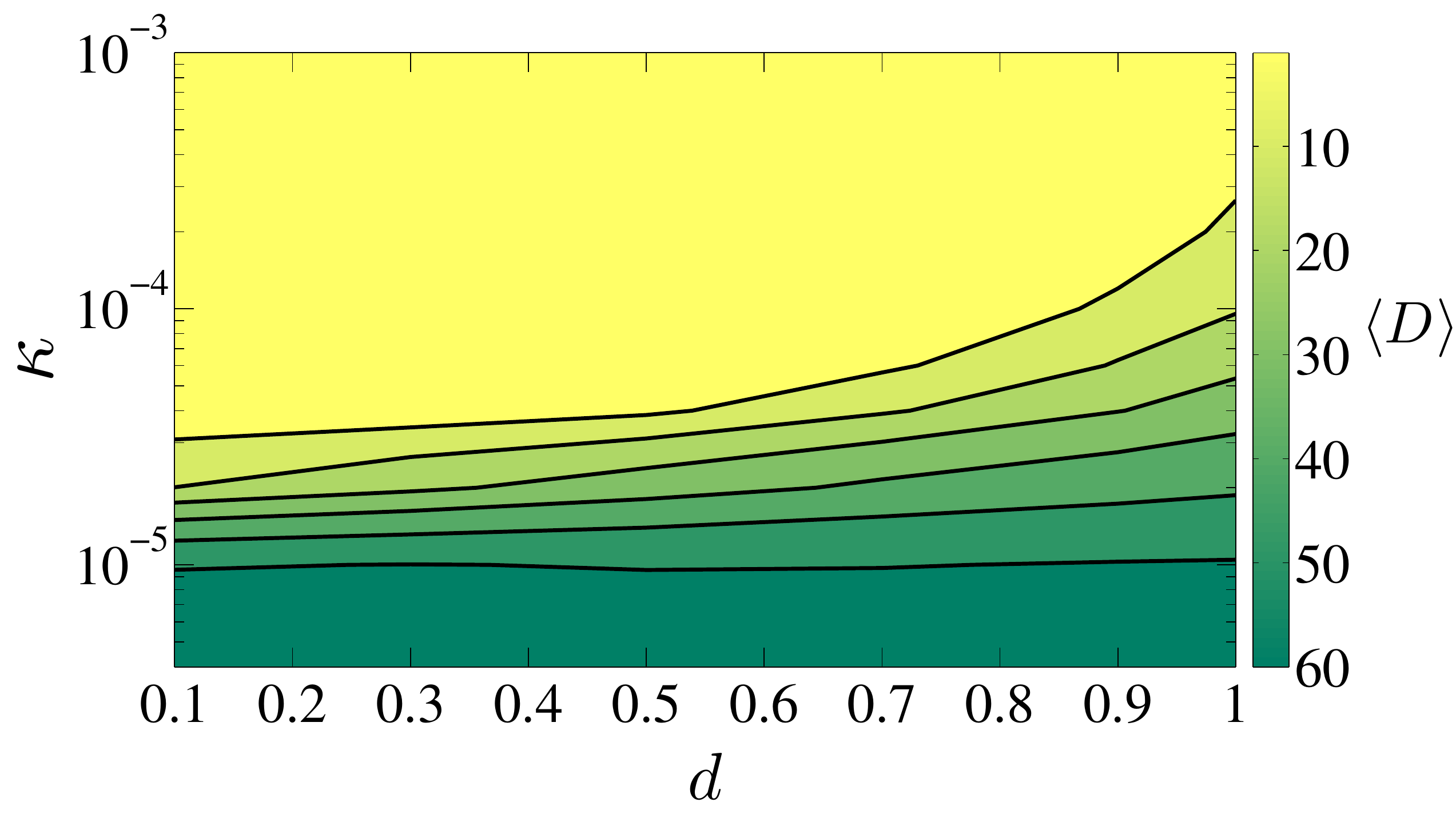}}
\caption{Contour-plot of the surface $\langle D(\kappa, d) \rangle$.
The system at time $t = 5 \times 10^5$; $\beta = 0$.}
\label{phasediagram}
\end{figure}
%


\subsection{Diffusivities leading to the competition success} \label{Sec-competition}



\subsubsection{Small temporal fluctuations ($\beta \ll \gamma$)} \label{Sec-small}


Because the initial diversity of the system is much larger than the final diversity,
the question about the competition advantage in the natural selection arises.
In fact, the high initial diversity was set for the purpose to see if some values of $\kappa_j$ can enhance the survival probability.
In order to answer the question, we study the probability distribution of $\kappa_j$ in the final state when $D = 1$ or at final simulation 
time $t_\mathrm{max}$.
In the first case, the final probability distribution is constructed using the global diffusion coefficients (in single realizations all 
individuals have finally the same diffusion coefficient).
In the second case, the different diffusion coefficients present in the systems at time $t_\mathrm{max}$ are used.

Figures~\ref{histogram}a and \ref{histogram-beta}a reveal that for small values of $\beta$ and $\kappa$, e.g., for $\beta = 0, \, 0.01$, 
and for $\kappa = 10^{-5}$, corresponding to the situation when each value of $\kappa_j$ leads to the strong clustering, the system is not 
very selective in diffusion coefficients and the final distribution of $P(\kappa_j)$ is rather similar to the initial (uniform) distribution.
Depending on temporal fluctuations, slower (small temporal fluctuations) or faster (large temporal fluctuations) diffusing bugs tend to be 
favored (see Fig.~\ref{histogram-beta}a). 
We remind that when increasing the temporal fluctuations of the system by increasing the value of $\beta$, we keep $\gamma = \mathrm{const}$, 
in order to have a comparable situation respect to the case $\beta = 0$, i.e., the critical number of neighbors, $N_R^*$, of particle $i$ is 
kept constant, see Eq.~(\ref{NR*}) and Sec.~\ref{Sec-model}.  
The reason why the initial and final probability distributions do not differ much lies in the fact that the variation in diffusivities is 
small and the probability to traverse the death-zones between the clusters, where the competition on resources is much higher than inside the 
clusters \cite{EHG-2015}, is very low for all values of $\kappa_j$.
Thus, the outcome of the competition is determined mostly by random fluctuations.
The small advantage given by the smaller diffusion coefficients when the temporal fluctuations are small (Fig.~\ref{histogram}a), is related 
primarily to the faster density enhancement near the individuals with smaller $\kappa_j$  when cluster formation takes place \cite{EHG-2015}.

When temporal fluctuations are small, the system is not very selective neither for large values of $\kappa$, as can be seen from 
Fig.~\ref{histogram}c for $\kappa = 10^{-3}$. 
The selection will take place when increasing $\beta$ and the competition advantage will be given to the faster diffusing individuals 
(see Fig.~\ref{histogram-beta}c).

Instead, the system is very selective in the diffusivities for intermediate values of $\kappa$ (e.g., for $\kappa = 10^{-4}$) as long as 
temporal fluctuations are sufficiently small.
The smaller the temporal fluctuations (the smaller is $\beta$), the more selective the system is (see Fig.~\ref{histogram-beta}b).
Thus, the influence of $\beta$ on the system selectiveness is the opposite in the cases of small and large and in the case of intermediate 
values of $\kappa$. 

For intermediate values of $\kappa$ the selectiveness of the system is influenced also by the value of $d$ (see Fig.~\ref{histogram}b).
If $d$ is sufficiently small, directional selection occurs in the system, i.e., the smallest diffusivities are favored.
Namely, with respect to the individuals with a larger flux out of the clusters, i.e., with larger diffusivities, the individuals diffusing 
slower and forming stronger clusters that are more compact and contain more individuals, experience less the high competition occurring 
between the clusters and have thus a higher probability for surviving \cite{Heinsalu-2013-PRL,EHG-2015}.
The directional selection is captured also through the mean-field approximation.

For larger values of $d$ a stabilizing selection not predicted by the mean-field theory 
\cite{Heinsalu-2013-PRL,EHG-2015,Hastings-1983,Dockery-1998} will take place, meaning that $P(\kappa_j)$ presents at $t_\mathrm{max}$ a clear 
maximum at smaller but intermediate values of $\kappa_j$, going then rather rapidly to zero; the distribution has a finite value at 
$\kappa_j \to 0$.
This situation is illustrated also by Fig.~\ref{clustering}d where one can notice that the clusters have different widths: the more compact 
clusters consist of slowly diffusing individuals and more spread clusters of more motile organisms.
The reason for such probability distribution shape will be discussed in Secs.~\ref{Sec-evolution} and \ref{Sec-1D} 
(see also Ref.~\cite{Heinsalu-2013-PRL}).
The distribution gets broader and the maximum shifts to larger values of $\kappa_j$ when increasing $\beta$, giving the competition advantage 
to faster diffusing individuals and leading to the disappearance of the slowly diffusing individuals from the system 
(see Fig.~\ref{histogram-beta}b).

From Figs.~\ref{histogram}a and \ref{histogram}c one can also estimate that in the case of small temporal fluctuations, for small and large 
values of $\kappa$ the mean diffusion coefficient of the system at later times will be approximately equal to $\kappa$ for any value of $d$.
Instead, for intermediate values of $\kappa$  the mean diffusion coefficient of the system at later times is smaller than $\kappa$ and 
depends on the value of $d$, being the smaller the smaller is $d$ (see Fig.~\ref{histogram}b).

\begin{figure}[!t]
\resizebox{0.45\textwidth}{!}{\includegraphics{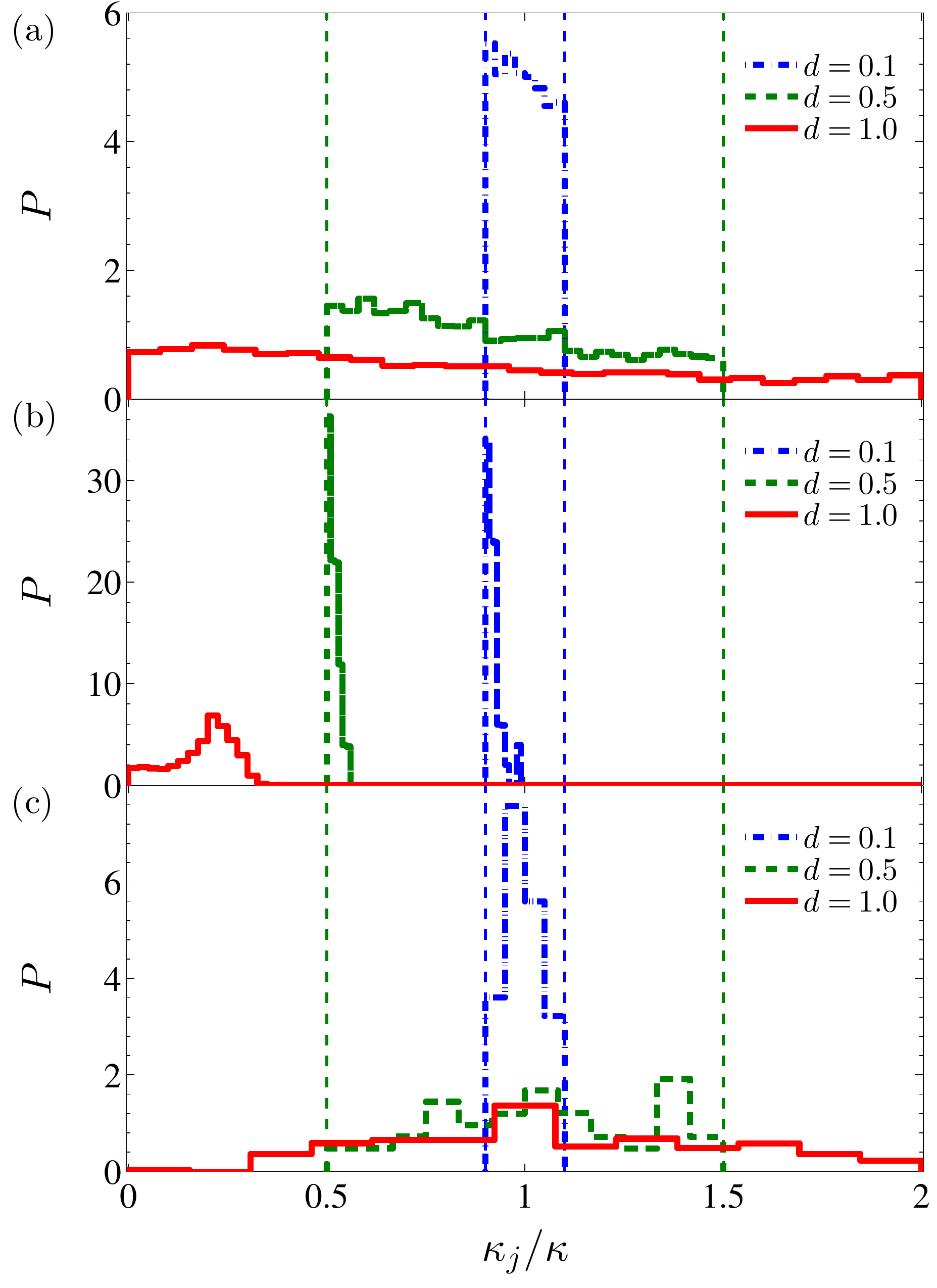}}
\caption{The probability distribution of diffusivities $\kappa_j$  for various values of $d$ and $\kappa$: (a) $\kappa = 10^{-5}$; (b) 
$\kappa = 10^{-4}$; (c) $\kappa = 10^{-3}$. The dashed vertical lines indicate the limits of the initial distributions; $\beta = 0$.}
\label{histogram}
\end{figure}
\begin{figure}[!t]
\resizebox{0.45\textwidth}{!}{\includegraphics{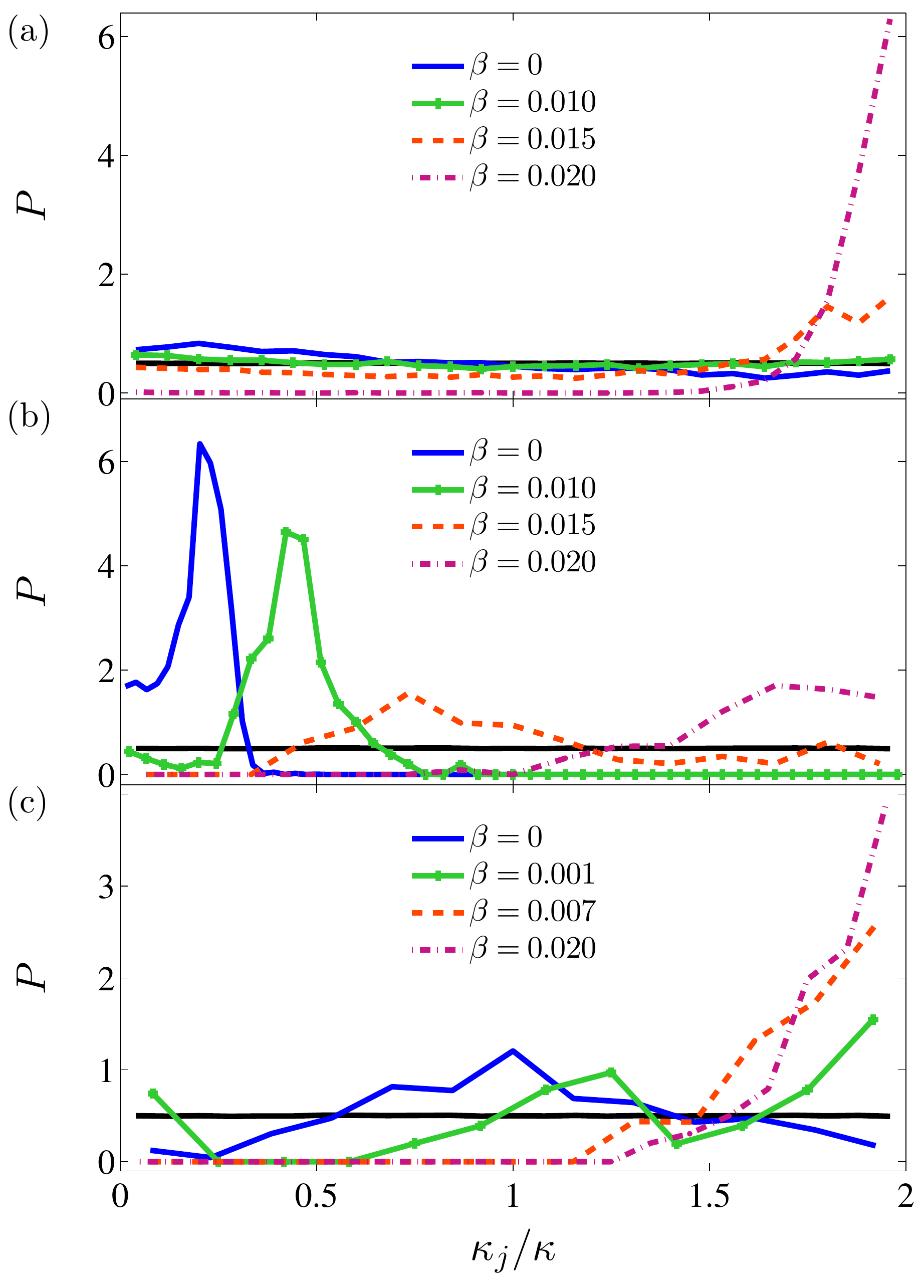}}
\caption{The probability distributions of diffusivities $\kappa_j$  for various values of $\beta$ 
and $\kappa$: (a) $\kappa = 10^{-5}$; (b) $\kappa = 10^{-4}$; (c) $\kappa = 10^{-3}$. The solid black lines represent the initial 
distributions ($d = 1$).}
\label{histogram-beta}
\end{figure}
%


\subsubsection{Influence of large temporal fluctuations} \label{Sec-large}


Figure~\ref{histogram-beta} reveals that when temporal fluctuations increase, the competition advantage is always given to the larger values 
of $\kappa_j$.
The reason is that larger fluctuations caused by the increase of the death rates lead to the disappearance of whole clusters, i.e., to the 
appearances of the empty spaces with favorable reproduction conditions and the organisms with larger diffusivities are more effective in 
occupying them (see also Refs.~\cite{Pigolotti-2014-PRL,Kessler-2009,Waddell-2010,Novak-2014,Johnson-1990,Lin-2015a}).
The same would happen keeping $\beta = 0$ but increasing $r_{d0}$ with $\Delta_0 = \mathrm{const}$.

As shown by Fig.~\ref{histogram-beta}c, for large $\kappa$ already a small increase of $\beta$ gives a strong competition advantage to the 
largest diffusivities. In the case of smaller values of $\kappa$ such transition is less prompt and for intermediate values of $\kappa$ one 
can observe a smooth shifting of the maximum of $P(\kappa_j)$ to larger values (the increase of temporal fluctuations is associated also to 
the increase of the critical diffusion coefficient leading to the clustering of 
organisms).


\subsection{Time evolution of the system} \label{Sec-evolution}


Let us analyze the time evolution  of the average total population size (average number of organisms), $\langle N \rangle$, and of the average 
diversity (average number of different diffusion coefficients present in the system), $\langle D \rangle$ (Fig.~\ref{evolution}).

Because initially the distribution of organisms is homogeneous and the population size $N_0$ very large, the number of individuals and the 
initial high diversity will decrease rapidly due to the competitive interactions until the balance between deaths and reproductions will be 
reached. Using Eq.~(\ref{NR*}) one can write for the corresponding particle density,
\begin{equation}
\label{density}
\rho^* = N_R^*/(\pi R^2)  \, ;
\end{equation}
the corresponding population size is $N^* \approx 1433$.
This state is reached the faster the larger is $\beta$,
i.e., the larger is the probability of death for $N_R^i > 0$ (Fig.~\ref{evolution}). 

For large values of $\kappa$ the organisms will be almost homogeneously distributed for any value of $d$ also at later times.
Consequently, the number of individuals will fluctuate around the value $N^*$. 
The diversity will continue to decrease until $D = 1$ due to the fluctuations and irreversibility of death.
We illustrate this in Fig.~\ref{evolution}c with the time evolution for the ensemble-averaged quantities for various values of $\beta$ and $d$.
As one can see, the decrease of the diversity is practically independent of the value of $d$.

Notice that the estimation (\ref{density}) is correct only for sufficiently large $\Delta_0$.
For small $\Delta_0$ the expected density is small and fluctuations in particle number will bring the system into stochastic extinction 
\cite{EHG-2004,CL-2004,EHG-2015}. 

For smaller values of $\kappa$ an increase of the population size follows after reaching $N^*$ (see Figs.~\ref{evolution}a and 
\ref{evolution}b), caused by the cluster formation.
In this regime the competition advantage is given to the slowly diffusing individuals for any value of $\beta$ due to the faster density 
enhancement near the bugs with smaller $\kappa_j$ \cite{EHG-2015}.

In the systems where all the individuals are identical, after the stationary state is reached, $\langle N(t) \rangle$ will fluctuate around 
a constant value.
For fixed parameter values the population size is the larger, the smaller is the diffusion coefficient, as discussed in great detail in 
Ref.~\cite{EHG-2015}; the maximum possible cluster size is determined by Eq.~(\ref{NR*}) and is reached for $\kappa \to 0$.
In the system consisting of individuals characterized by different diffusivities, the average total population size, $\langle N(t) \rangle$, 
is determined by the average diffusion coefficient of the system, $\langle \kappa_j \rangle$, at time $t$.
Thus, until the system reaches the state with $\langle D \rangle = 1$ or $\langle D(t) \rangle = \mathrm{const.}$, the average total 
population size will be a time dependent quantity.  
Because large temporal fluctuations give the competition advantage to faster diffusing individuals, leading to the larger value of 
$\langle \kappa_j \rangle$ (Fig.~\ref{histogram-beta}), the increase of $\beta$ leads to the decrease of the average population size at 
time $t$ (difficult to see from Fig.~\ref{evolution}).

While at small and large values of $\kappa$ the time evolution of diversity is practically independent of $d$ (see Figs.~\ref{evolution}a 
and \ref{evolution}c), at intermediate values of $\kappa$ it is greatly influenced by it (Fig.~\ref{evolution}b).
After the decrease corresponding to the population decline due to the competition between the individuals caused by the high initial 
crowdedness, the decrease of the diversity continues until its value equals the number of clusters, $N_c$, that the hexagonal pattern 
appearing in the system can fit.
Until that moment the behavior of $\langle D(t) \rangle$ is practically independent of $d$ and smaller diffusivities that are more effective 
in cluster formation are more favorable.
Starting from $\langle D(t) \rangle \approx N_c$ the competition between the subpopulations occupying different clusters will take place. 
In the case of an intermediate value of $\kappa$ (Fig.~\ref{evolution}b), the further decrease of diversity is the slower the larger is the 
value of $d$ due to the larger fraction of the individuals with smaller diffusivities forming stronger clusters that are more resistant to 
the invaders (see also Sec.~\ref{Sec-RT}).
At the same time, the bugs with smaller diffusivities that are more successful in the intra-cluster competition are less successful in the 
inter-cluster competition.
Such interplay between the inter- and intra-cluster competition leads to the existence of the optimal diffusivity range observed in 
Figs.~\ref{histogram}b and \ref{histogram-beta}b (see also Sec.~\ref{Sec-1D}).
The situation is similar to the one observed in Ref.~\cite{Heinsalu-2013-PRL} for the coexistence of competing Brownian and L\'evy bugs.
The small probability to traverse successfully the inter-cluster space is also the reason why for small values of $\kappa$ the diversity 
turns at $\langle D \rangle \approx N_c$ on a plateau for any $d$, as can be seen from Fig.~\ref{evolution}a 
(see also Ref.~\cite{Heinsalu-2012-PRE})).

For $\beta > 0$ one can observe from Fig.~\ref{evolution}b that after reaching the value $D = N_c$ the decrease of $D$ slows down for a 
while, but speeds up again when $P(\kappa_j \to 0)$ starts to decrease due to the temporal fluctuations.

If instead of Brownian motion the organisms would perform L\'evy walks, the maximum in $P(\kappa_j)$ for intermediate values of $\kappa$ and 
the large diversity for small values of $\kappa$ would not be observed in the systems with small temporal fluctuations.
In this case, due to the occasional long jumps it would always be possible to arrive after some time to the other clusters and the outcome 
would be determined by the intra-cluster competition and the value $D = 1$ would be observed within a reasonable simulation 
time \cite{Heinsalu-2012-PRE}.
The same is valid when taking into account the mutation process.
In both cases it would be always the individuals with the smallest diffusion coefficients to win the competition 
\cite{Heinsalu-2013-PRL,EHG-2015}.
 
In Sec.~\ref{Sec-small} it was observed that for intermediate values of $\kappa$ the average diffusion coefficient of the system is the 
smaller the larger is $d$ (Fig.~\ref{histogram}b).
Thus, in the case of intermediate $\kappa$, the population size is the larger the larger is $d$ 
(difficult to see from Fig.~\ref{evolution}b).

\begin{figure}[!t]
\resizebox{0.45\textwidth}{!}{\includegraphics{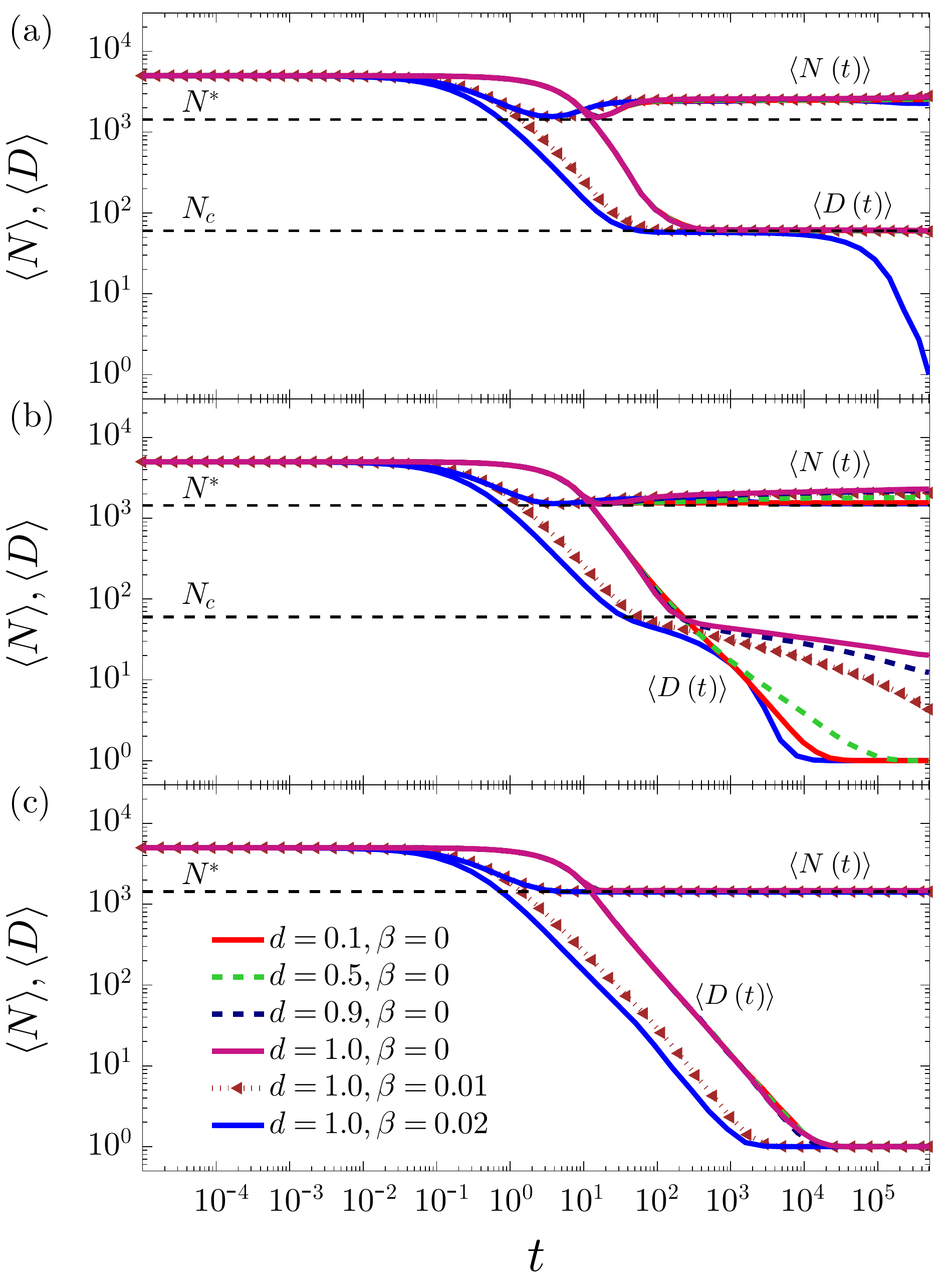}}
\caption{Time evolution of the average number of organisms, $\langle N \rangle$, and diversity, $\langle D \rangle$.
(a) $\kappa = 10^{-5}$; (b) $\kappa = 10^{-4}$; (c) $\kappa = 10^{3}$; different curves in the figures correspond to different 
values of $d$.}
\label{evolution}
\end{figure}
%


\subsection{Residence times in states with different diversities} \label{Sec-RT}


It is interesting to investigate also the residence times, $\tau$, in states with different diversities (Fig.~\ref{residence}).
Let us first investigate the case $\beta = 0$ (small temporal fluctuations).
From Figs.~\ref{residence}b and \ref{residence}c one can see that for intermediate and large values of $\kappa$ with $d$ such that the 
system will reach the value $D = 1$ within an accessible simulation time, there are two power-law regimes for the average residence times. 
For small and intermediate values of $\kappa$ with $d$ such that the state with $D = 1$ is not reached within an accessible simulation 
time $t_\mathrm{max}$, the behavior of residence times versus diversity is more complicated (Figs.~\ref{residence}a and \ref{residence}b).
In the interval $D=N_0$ until $D = N^*$ the average residence times follow a power-law with power equal to $1$ for any $\kappa$ and $d$ and 
it is associated with the decrease of the competition.
The second power-law depends on $\kappa$ (and for intermediate values of $\kappa$ slightly on $d$), being for large $\kappa$ equal to $2$ 
and decreasing with  decreasing $\kappa$.
Notice that when there is no strong clustering in the system and the state with $D = 1$ will be reached within an accessible simulation 
time, the residence times for $D\to 1$ grow even faster than the second power-law, meaning that the competition between the last species 
remained is hardest.
Instead, for intermediate values of $\kappa$ with $d$ leading to strong clustering, the residence times grow significantly already at 
$D \approx N_c$ in association with the competition between the species from different clusters, as discussed in Sec.~\ref{Sec-evolution}; 
the growth of the residence times is the stronger the larger is $d$, but seems to slow down a little with the decrease of diversity in the 
system.
For small values of $\kappa$ the residence times seem to diverge at $D \approx N_c$, increasing in the interval $D \approx 70$ to 
$D \approx 60$ from $\langle \tau \rangle \sim 10$ to $\langle \tau \rangle \sim 10^5$.
Thus, we can conclude that for accessible simulation times there is a finite diversity $D > 1$ in the system; from Fig.~\ref{residence}a one 
cannot even estimate after how long time the state $D=1$ could be reached.

The increase of temporal fluctuations leads to the decrease of residence times, i.e., the disappearance of different diffusivities 
(decrease of diversity) is the faster the larger is $\beta$ (see the curves for $\beta > 0$ in Fig.~\ref{residence}).
Consequently, also the cluster formation takes place at smaller times, as can be seen from Fig.~\ref{evolution}.

\begin{figure}[!t]
\resizebox{0.45\textwidth}{!}{\includegraphics{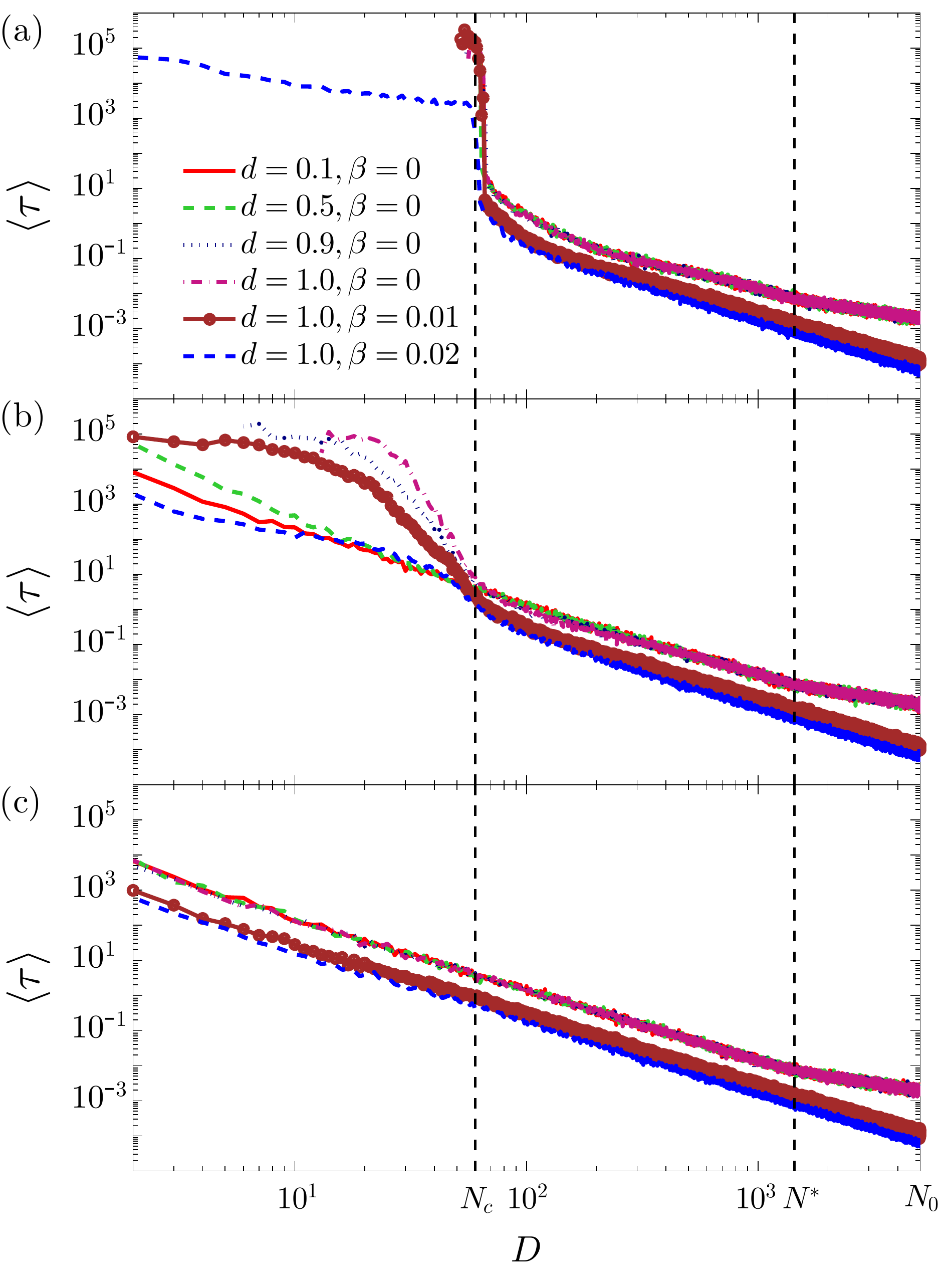}}
\caption{Average residence times $\langle \tau \rangle$ in states with different diversities $D$.
(a) $\kappa = 10^{-5}$; (b) $\kappa = 10^{-4}$; (c) $\kappa = 10^{3}$; different curves correspond to different values of $d$.
Notice that the systems evolve with time from right to left, i.e., $D = N_0$ corresponds to the initial time $t = 0$ and while the time 
passes the value of $D$ decreases.}
\label{residence}
\end{figure}
%


\subsection{Optimal diffusion in a one-dimensional system} \label{Sec-1D}


Figures~\ref{evolution}b and \ref{residence}b reveal that waiting a very long time the system will reach finally the state with $D=1$ also in 
the case $\beta =0$.
Thus, one can question the result depicted in Figs.~\ref{histogram}b and \ref{histogram-beta}b and discussed in Sec.~\ref{Sec-small} that 
stabilizing selection takes place and there exists an optimal diffusivity range leading to the increase of the competition success.
In fact, it is not predicted by the mean-field theory which describes only the directional selection 
\cite{Heinsalu-2013-PRL,EHG-2015,Hastings-1983,Dockery-1998}.
Thus, one could suspect that this is just an effect of the transition and finally it is still the family with the smallest diffusion 
coefficient that will win the competition.
To verify that this is not the case, we have investigated the one-dimensional system that converges to the final state with $D=1$ within an
accessible simulation time.
This allows us to make a comparison between numerical results and analytical estimates.
The result is depicted in Fig.~\ref{histogram-1dim}.
In this figure $\beta =0$, $\kappa = 10^{-4}$, and $d = 1$.
As one can observe, the distribution $P(\kappa_j)$ shows a similar behavior as the curves in Figs.~\ref{histogram}b and \ref{histogram-beta}b 
for the same parameters.
Thus, it is not a transient.

Following the discussion of Sec.~\ref{Sec-evolution} that the stabilizing selection should occur as a result of the interplay between the 
inter- and intra-cluster competition, the value of $\kappa_j$ corresponding to the maximum of the probability distribution can be calculated 
from the following condition:
\begin{equation}
\label{max}
t^* = t_m  \, .
\end{equation}
Here $t^*$ is the typical first-passage time of an organism with the diffusion coefficient $\kappa_j$ for diffusing from the center of one 
cluster to the center of the other cluster in a one-dimensional system,
\begin{equation}
\label{t*}
t^* = \delta^2/(6 \kappa_j)  \, .
\end{equation}
$\delta$ is the distance between the cluster centers and has been previously estimated from the mean-field theory to be 
$\delta \approx 0.131475$ \cite{EHG-2015,EHG-2005}.
Instead, $t_m$ is the typical lifetime of a family defined as a chosen individual and its descendants.
Thus, even if an individual with diffusivity $\kappa_j$ does not reach a neighboring cluster, its descendants that continue the diffusion 
process of the mother may arrive there and in this way the organisms with a certain $\kappa_j$ can invade new clusters even if $\kappa_j$ 
is rather small (so that the probability for a single organism to arrive there is extremely small).
Following Ref.~\cite{EHG-2005},
\begin{equation}
\label{tm}
t_m = \Delta_0/(2 \alpha r_{d0})  \, .
\end{equation}
Thus, from conditions~(\ref{max}), (\ref{t*}), (\ref{tm}) we get that the optimal diffusion coefficient is
\begin{equation}
\label{kappaopt}
{\kappa_j}^* = \delta^2 \alpha r_{d0} / (3 \Delta_0)  \, .
\end{equation}
For the parameters used in Fig.~\ref{histogram-1dim} we have that ${\kappa_j}^* = 0.128 \times 10^{-4}$, which agrees well with the numerical 
result.
Organisms (families) with smaller diffusion coefficient most probably do not reach other clusters and organisms with larger diffusivities 
are not successful in the intra-cluster competition.

\begin{figure}[!t]
\resizebox{0.45\textwidth}{!}{\includegraphics{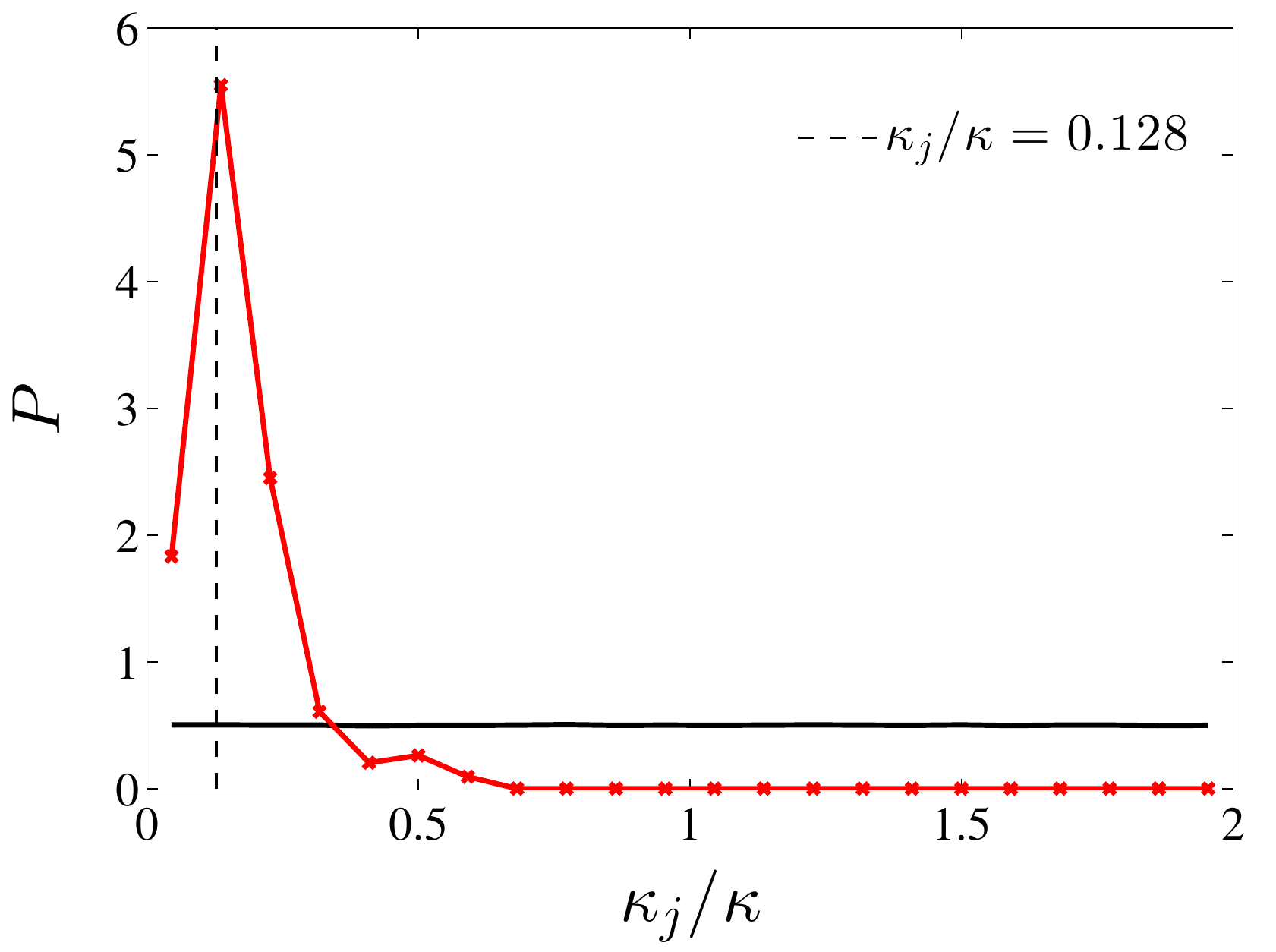}}
\caption{The probability distributions of diffusivities $\kappa_j$ in the final state with $D = 1$ in a one-dimensional system for 
$\beta =0$ 
and $\kappa = 10^{-4}$. The solid black line represents the initial distributions ($d = 1$). The vertical dashed line represents the 
theoretical result from Eq.~(\ref{kappaopt}).}
\label{histogram-1dim}
\end{figure}
%


\section{Conclusion} \label{Sec-conclusion}


As mentioned in the Introduction, the problem addressed in the present paper corresponds to the process of natural selection. 
At the same time, in a certain parameter range, the model could be seen also as a model of competition and parapatric speciation.
Namely, the original population consists of $N_0$ individuals that have different diffusivities $\kappa_j$, with $j = 1, \dots, N_0$, i.e., 
there is a variance in their foraging behaviors, but are identical in all the rest.
Due to the competitive interactions, asexual reproduction, and reproductive correlations, for a certain range of parameters, the bugs will be 
separated after some time into different zones, each occupied by bugs coming from a single ancestor.
Because such separated subpopulations have different behaviors, we can define them as different species.  
The competition keeps the subpopulations in the separated zones approximately constant for the chosen parameter values and also the total 
population size of the system is approximately constant, fluctuating only weakly around the average value.
Because the inter-cluster traveling is possible then the individuals of each species may come in contact or cross habitats from time to time 
and the invasion of new territories is also possible, i.e., a species can occupy more than one separated zone.

Investigating mostly numerically the individual-based model we have observed that the mean value $\kappa$ and the relative width $d$ of the 
initial distribution of the diffusion coefficients together with the temporal fluctuations determine the final distribution of the 
diffusivities (diffusion coefficients leading to the competition success) as well as the final diversity of the system at finite time 
(the number of different diffusion coefficients present in the system).
We have seen that large initial mean diffusivity of a system leads to a rather fast disappearance of the diversity whereas a small initial 
mean diffusivity leads to the diversity equal to the number of the self-organized patches of the individuals.
It is also shown that depending on the parameter values, introducing the heterogeneity in diffusion coefficients can lead to the clustering 
of individuals that in turn leads to the slower disappearance of the diversity. 
Thus, the diversity is related to the spatial heterogeneities (patch formation) --- as known well from previous studies --- and the resulting 
inter-cluster competition.
The clustering of the organisms is also associated to the enhanced competition success of the slower diffusing individuals 
\cite{Heinsalu-2013-PRL,EHG-2015,Hastings-1983,Holt-1985,Dockery-1998,Lin-2015b,Hutson-2003,Dieckmann-1999,Hutson-2001,Baskett-2007}.
The diversity is diminished by the increase of the temporal fluctuations that give the competition advantage to the faster diffusing 
individuals \cite{Pigolotti-2014-PRL,Kessler-2009,Waddell-2010,Novak-2014,Johnson-1990,Lin-2015a}.
However, for a large parameter range we have shown that the extreme values of the diffusivities do not lead to the largest competition 
success.
Instead, in most cases there exists an optimal range of diffusion coefficients giving the competition advantage.
This result is rather uncexpected and is not captured in the mean-field theory.
As we have demonstrated, in the case of a one-dimensional system that allows the comparison between the numerical and analytical 
calculations, the observed stabilizing selection occurs as a result of the interplay between the inter- and intra-cluster competition.


Finally, let us mention that though we have not assumed that a larger dispersal has a cost, this feature emerges naturally in populated 
environments, where spatial inhomogeneities occur (in the present case due to the reproductive correlations) and the temporal fluctuations 
are not too large, due to the neighborhood dependent reproduction rates (\ref{rates-b}).
Namely, for $\alpha > 0$, the larger is the diffusion coefficient of an individual the lower is on average its reproduction rate; the effect 
is the larger the stronger is the clustering (see the discussion in Ref.~\cite{EHG-2015}).
Furthermore, in the case of small temporal fluctuations, considering the sexual reproduction and taking into account the Allee effect does 
not make any difference in the results because the individuals gathered in clusters have many neighbors and the ones between the clusters even 
more. The Allee effect plays a role only in the case of large temporal fluctuations when there are empty space regions forming due to the 
disappearance of the clusters.\\


\subsection*{Acknowledgments} \label{Sec-conclusion}


This work was supported by institutional research funding IUT (IUT-39) and Personal Research Funding Grant PUT (PUT-1356) of the Estonian 
Ministry of Education and Research and EU through the European Regional Development Fund (ERDF) Center of Excellence (CoE) program grant 
TK133 .
E.H. wants to thank E. Hern\'andez-Garc\'ia for useful comments.

\subsection*{Authors contributions statement}
All the authors were involved in writing the code. D.N.M. performed the numerical simulations and data analysis. E.H. made the analytical 
calculations. All the authors were involved in the preparation of the manuscript. All the authors have read and approved the final manuscript.

%


\end{document}